\documentclass[12pt]{article}
\usepackage{amssymb,latexsym,amsmath}
\begin{document}

\title{\textbf{De Sitter Universe in Nonlocal Gravity}}

\author{E. Elizalde$^{1}$\footnote{E-mail: elizalde@ieec.uab.es, elizalde@math.mit.edu} ,
E.O. Pozdeeva$^{2}$\footnote{E-mail: pozdeeva@www-hep.sinp.msu.ru} , and  S.Yu.
Vernov$^{1,2}$\footnote{E-mail: vernov@ieec.uab.es, svernov@theory.sinp.msu.ru} \vspace*{3mm} \\
\small $^1$Instituto de Ciencias del Espacio (ICE/CSIC) \, and \\
\small  Institut d'Estudis Espacials de Catalunya (IEEC), \\
\small  Campus UAB, Facultat de Ci\`encies, Torre C5-Parell-2a planta, \\
\small  E-08193, Bellaterra (Barcelona), Spain\\
\small  $^2$Skobeltsyn Institute of Nuclear Physics,  Lomonosov Moscow State University,\\
\small  Leninskie Gory 1, 119991, Moscow, Russia
}

\date{ }

\maketitle


\vspace{-7.2mm}
\begin{abstract}
A nonlocal gravity model,  which does not assume the existence of a new dimensional parameter in the action and includes a function $f(\Box^{-1} R)$, with $\Box$ the d'Alembertian operator,  is considered. The model is proven to have de Sitter solutions only if the function $f$ satisfies a certain second-order linear differential equation. The de Sitter solutions corresponding
to the simplest case, an exponential function $f$, are explored, without any restrictions on the parameters. If the value of the Hubble parameter is positive, the de Sitter solution is stable
at late times, both for negative and for positive values of the cosmological constant. Also, the
stability of the solutions with zero cosmological constant is discussed and
sufficient conditions for it are established in this case. New de Sitter solutions
are obtained which correspond to the model with dark matter and
stability is proven in this situation for nonzero values of the cosmological constant.
\end{abstract}



\section{Introduction}

Modern cosmological observations, such as those coming from Supernovae Ia (SNe
Ia)~\cite{SN1}, from the cosmic microwave background (CMB) radiation~\cite{WMAP,
Komatsu:2010fb}, large scale structure (LSS)~\cite{LSS}, baryon
acoustic oscillations (BAO)~\cite{Eisenstein:2005su}, and weak
lensing~\cite{Jain:2003tba}, allow to obtain joint constraints on cosmological parameters
(see, for example,~\cite{Kilbinger:2008gk})
and indicate that the current expansion of the
Universe is accelerating. There are a few types of models able to reproduce
this late-time cosmic acceleration. The simplest one is general
relativity with a cosmological constant (for a review, see,
e.g.,~\cite{Weinberg:1988cp}). Some others involve modifications of
gravity, as for instance $F(R)$ gravity, with $F(R)$ an (in principle) arbitrary
function of the scalar curvature $R$ (for recent reviews see,
e.g.,~\cite{Review-Nojiri-Odintsov,Book-Capozziello-Faraoni}).

Modified gravity cosmological models have been proposed in the hope of
finding solutions to the important open problems of the standard cosmological
model. There are lots of ways to deviate from Einstein's gravity.
Different modifications of this theory have been considered in detail in the
reviews~\cite{Review-Nojiri-Odintsov,Nojiri:2006ri}.
As a promising modification of gravity, the nonlocal gravity theory obtained by
taking into account quantum effects has been proposed in~\cite{Deser:2007jk}. Also,
as is well known string/M-theory is usually considered as a possible theory for all
fundamental interactions, including gravity; again, the appearance of
nonlocality within string field theory is a good motivation for
studying nonlocal cosmological models. Moreover,
there was a proposal on the solution of the cosmological constant
problem by a nonlocal modification of gravity~\cite{ArkaniHamed:2002fu}.
 The majority of nonlocal cosmological models explicitly include a
function of the d'Alembert operator, $\Box$, and either define a
nonlocal modified
gravity~\cite{Non-local-gravity-Refs,Odintsov0708,Jhingan:2008ym,Koivisto:2008,Nojiri:2010pw,
1104.2692,Farajollahi,ZS,Non-local-FR}
or add a nonlocal scalar field, minimally coupled to
gravity~\cite{Non-local_scalar}.

In this paper, we consider a modification that includes a function of
the $\Box^{-1}$ operator. Such modification does not assume the
existence of a new dimensional parameter in the action. This nonlocal
model has a local scalar-tensor formulation. A modification of
nonlocal gravity with a  term $f(\Box^{-1} R)$ has been studied in
order to realize a unified scenario, comprising both early-time
inflation and the late-time cosmic acceleration. It has been shown
in~\cite{Odintsov0708} that a theory of this kind, being consistent
with Solar System tests, may actually lead to the known Universe
history sequence: inflation, radiation/matter dominance and a dark
epoch. An explicit mechanism to screen the cosmological constant in
nonlocal gravity was discussed in~\cite{Nojiri:2010pw,1104.2692,ZS}.
Different cosmological aspects of such nonlocal gravity models have been studied
in~\cite{Jhingan:2008ym,Koivisto:2008,Nojiri:2010pw,1104.2692,Farajollahi,ZS,Non-local-FR}, too.

In this paper, we explore in detail de Sitter solutions in the nonlocal gravity
model. We prove that the model can have de Sitter solutions only if the
function $f(\Box^{-1}R)$ satisfies a certain second order linear differential
equation, which is a nice result.
The simplest and most studied model~\cite{Odintsov0708,Jhingan:2008ym,Koivisto:2008,Nojiri:2010pw,1104.2692,Farajollahi,ZS} admitting de Sitter solutions
 is characterized by a
function $f(\Box^{-1}R)=f_0 e^{(\Box^{-1}\!R)/\beta}$, where $f_0$ and
$\beta$ are real parameters.
A few de Sitter solutions for this model have been found
in~\cite{Odintsov0708} and also analyzed in~\cite{1104.2692}. In both
papers the authors put strict restrictions on arbitrary parameters
(integration constants).

Here, we will consider all possibilities for de Sitter solutions
without any restriction. We will also obtain de Sitter solutions in the
case when the matter included in the model is dark matter. Finally, we
will analyze the stability of these de Sitter solutions and get the
corresponding restrictions on the parameters of the model.
In the case $\Lambda=0$, the system of equation, describing this model,
has been written in terms of Hubble-normalized variables and the
stability of the fixed point of this system has been
analyzed in~\cite{Odintsov0708}. This has been continued in~\cite{Jhingan:2008ym}
(see also~\cite{Farajollahi}). We will consider this specific case in
Subsection~\ref{4-2}.

The paper is organized as follows. In Section~2, we shortly review
nonlocal gravity models that have a local scalar-tensor formulation.
In Section~3 we obtain the necessary conditions on the function $f(\Box^{-1}R)$,
which allow us to get de Sitter solutions, and then look for general de
Sitter solutions in the case of the exponential function
$f(\Box^{-1}R)$. In Section~4 we discuss the stability of the de Sitter
solutions for Friedmann--Lema\^{i}tre--Robertson--Walker (FLRW) and
Bianchi I metrics. Section~5 is devoted to conclusions.

\section{Nonlocal gravitational model}
Consider the following action for nonlocal gravity
\begin{equation}
\label{nl1}
S=\int d^4 x \sqrt{-g}\left\{
\frac{1}{2\kappa^2}\left[ R\left(1 + f(\Box^{-1}R )\right) -2 \Lambda \right]
+ \mathcal{L}_\mathrm{matter}
\right\}
\, ,
\end{equation}
where ${\kappa}^2 \equiv 8\pi/{M_{\mathrm{Pl}}}^2$, the Planck mass
being $M_{\mathrm{Pl}} = G^{-1/2} = 1.2 \times 10^{19}$ GeV. We use the
signature $(-,+,+,+)$, $g$ is the determinant of the metric tensor
$g_{\mu\nu}$, $f$  a differentiable function,  $\Lambda$  the
cosmological constant, and $\mathcal{L}_\mathrm{matter}$ is the matter
Lagrangian. Recall the covariant d'Alembertian for a scalar field, which reads
\begin{equation}
\label{BOX} \Box \equiv \frac{1}{\sqrt{-g}} \partial_\mu \left(
\sqrt{-g} \, g^{\mu
  \nu}\partial_\nu \right).
\end{equation}
Introducing two scalar fields, $\eta$ and $\xi$, we can rewrite
action~(\ref{nl1}) in the following local form:
\begin{equation}
\label{anl2}
S =
\int d^4 x \sqrt{-g}\left\{
\frac{1}{2\kappa^2}\left[R\left(1 + f(\eta)-\xi\right)
+ \xi\Box\eta  - 2 \Lambda \right]
+ \mathcal{L}_\mathrm{matter}  \right\}
\, .
\end{equation}
By varying the action (\ref{anl2}) over $\xi$, we get
$\Box\eta=R$.
Substituting $\eta=\Box^{-1}R$ into action~(\ref{anl2}), one reobtains
action~(\ref{nl1}).
Varying action~(\ref{anl2}) with respect to the metric tensor
$g_{\mu\nu}$, one gets
\begin{equation}
\label{nl4}
\begin{split}
&\frac{1}{2}g_{\mu\nu} \left[R\left(1 + f(\eta) -
 \xi\right)
 - \partial_\rho \xi \partial^\rho \eta - 2 \Lambda \right]
 - R_{\mu\nu}\left(1 + f(\eta) - \xi\right)+\\ &+ \frac{1}{2}\left(\partial_\mu \xi \partial_\nu \eta
+ \partial_\mu \eta \partial_\nu \xi \right)
 -\left(g_{\mu\nu}\Box - \nabla_\mu \partial_\nu\right)\left( f(\eta) -
\xi\right)
+ \kappa^2T_{\mathrm{matter}\, \mu\nu}=0\, ,
\end{split}
\end{equation}
where $\nabla_\mu$ is the covariant derivative and
$T_{\mathrm{matter}\,\mu\nu}$ the energy--momentum tensor of matter,
defined as
\begin{equation}\label{Tmatter}
T_{\mathrm{matter}\, \mu\nu} \equiv
-\frac{2}{\sqrt{-g}}
 \frac{\delta \sqrt{-g} {\mathcal{L}}_{\mathrm{matter}}}{\delta g^{\mu \nu}}.
\end{equation}
On the other hand, variation of  action~(\ref{anl2}) with respect to $\eta$ yields
\begin{equation}
\label{nl5}
\Box\xi+ f'(\eta) R=0 ,
\end{equation}
where the prime denotes derivative with respect to $\eta$.

We take the spatially flat FLRW metric,
\begin{equation}
\label{mFr}
ds^2={}-dt^2+a^2(t)\left(dx_1^2+dx_2^2+dx_3^2\right),
\end{equation}
and consider the case where the scalar fields $\eta$ and $\xi$
depend on time only. In the FLRW metric, system of Eqs.~(\ref{nl4}) is
equivalent to  the following equations:
\begin{equation}
\label{equ1}
{}- 3 H^2\left(1 + f(\eta) - \xi\right) + \frac{1}{2}\dot\xi \dot\eta
 - 3H\frac{d}{dt}  \left( f(\eta) -
\xi \right)  + \Lambda
+ \kappa^2 \rho_{\mathrm{m}}=0\, ,
\end{equation}
\begin{equation}
\label{equ2}
\left(2\dot H + 3H^2\right) \left(1 + f(\eta) - \xi\right)
+ \frac{1}{2}\dot\xi \dot\eta
+ \left(\frac{d^2}{dt^2} + 2H \frac{d}{dt} \right) \left( f(\eta) -
\xi \right) - \Lambda + \kappa^2 P_{\mathrm{m}}=0\, ,
\end{equation}
where $H=\dot{a}/a$ is the Hubble parameter, the dot denoting time
derivative. For a perfect matter fluid, we have $T_{\mathrm{matter}\, 0 0} =
\rho_{\mathrm{m}}$ and $T_{\mathrm{matter}\, i j} = P_{\mathrm{m}} g_{i
j}$. The  equation of state (EoS) is
\begin{equation}
\label{equ_rho} \dot\rho_{\mathrm{m}}={}-
3H(P_{\mathrm{m}}+\rho_{\mathrm{m}}).
\end{equation}
Adding up Eqs.~(\ref{equ1}) and (\ref{equ2}), we get
\begin{equation}
\label{equ12} 2\dot H \left(1 + f(\eta) - \xi\right) + \dot\xi \dot\eta
+ \left(\frac{d^2}{dt^2} - H \frac{d}{dt} \right) \left( f(\eta) - \xi
\right) + \kappa^2 (P_{\mathrm{m}}+\rho_{\mathrm{m}})=0 .
\end{equation}
Furthermore, from $\Box \eta = R$ and (\ref{nl5})
the equations of motion for the scalar fields $\eta$ and $\xi$ follow
\begin{equation}
\label{equ3} \ddot \eta + 3H \dot \eta ={}- 6 \left(\dot H + 2
H^2\right) ,
\end{equation}
\begin{equation}
\label{equ4}
\ddot \xi + 3H \dot \xi =6\left( \dot H + 2 H^2\right)f'(\eta) \, ,
\end{equation}
where we have used $R = 6\dot{H} + 12H^2$.

Let us consider the system of equations (\ref{equ_rho})--(\ref{equ4}).
Together with (\ref{equ1}), they are equivalent to the full
system of Einstein's equations. Differentiating (\ref{equ1}) with respect to $t$
and substituting into (\ref{equ2}), (\ref{equ_rho}), (\ref{equ3}), and
(\ref{equ4}), we get that (\ref{equ1}) is an integral of motion for the
system of equations (\ref{equ_rho})--(\ref{equ4}). Therefore, to find a
solution of the Einstein equation one can solve the system
(\ref{equ_rho})--(\ref{equ4}), which does not include the cosmological
constant $\Lambda$. After that, substituting into (\ref{equ1}) the
initial values of the solution obtained, one gets the corresponding
value of $\Lambda$.

The system of equations considered does not include the function
$\eta$, but only $f(\eta)$, $f'(\eta)$ and time derivatives of $\eta$. This
property can be used to analyze the stability of the de Sitter
solutions.

\section{De Sitter solutions}
\subsection{Nonlocal models with de Sitter solutions}

We now assume that the Hubble parameter is a nonzero constant: $H =
H_0$. In this case, Eq.~(\ref{equ3}) has the following general solution:
\begin{equation}
\label{eta} \eta(t) = {}-4H_0(t-t_0) - \eta_0e^{-3H_0(t-t_0)},
\end{equation}
with integration constants $t_0$ and $\eta_0$. All equations are
homogeneous, so if a de Sitter solution exists at $t_0=0$, then it
exists at an arbitrary $t_0$. So, without loss of generality we can set
$t_0=0$.

Subtracting Eq.~(\ref{equ1}) from Eq.~(\ref{equ2}), we get a linear
differential equation
\begin{equation}
\label{equPsi}
\ddot\Psi+5H_0\dot\Psi+6H_0^2(1+\Psi)-2\Lambda+\kappa^2(w_{\mathrm{m}}-1)\rho_{\mathrm{m}}=0,
\end{equation}
where $\Psi(t)=f(\eta(t))-\xi(t)$.

Solving (\ref{equPsi}) and substituting $\xi(t)=f(\eta(t))-\Psi(t)$
into Eq.~(\ref{equ4}), we get a linear differential equation for
$f(\eta)$
\begin{equation}
\label{equaf}
\dot\eta^2 f''(\eta)+\left(\ddot\eta+3H_0\dot\eta-12H_0^2\right)f'(\eta)=\ddot\Psi+3H_0\dot\Psi\,.
\end{equation}
Therefore, the model, which is described by action~(\ref{anl2}), can
have de Sitter solutions only if $f(\eta)$ satisfies Eq.~(\ref{equaf}).
In other words Eq.~(\ref{equaf}) is a necessary condition that the
model has de Sitter solutions. To prove the existence of de Sitter
solutions for the given $f(\eta)$ one should also check Eqs.
(\ref{equ1}) and (\ref{equ2}). Note that Eq.~(\ref{equaf}) has been
obtained without any restrictions on solutions and the perfect matter
fluid.

To demonstrate how one can get $f(\eta)$, which admits the existence of
de Sitter solutions, in the explicit form, we restrict ourselves to the
case $\eta_0=0$. In this case, Eq.~(\ref{equaf}) has the following form:
\begin{equation}
\label{equ7}
16H_0^2f''(\eta)-24H_0^2f'(\eta)=\Phi(\eta),
\end{equation}
where $\Phi(\eta)=\Phi(-4H_0t)\equiv \ddot\Psi+3H_0\dot\Psi$. We get
the following solution
\begin{equation}
f(\eta) =
\frac{1}{16H_0^2}\int\limits^\eta\left\{\int\limits^\zeta\Phi(\tilde{\zeta})e^{-3\tilde{\zeta}/2}d\tilde{\zeta}
+16C_3H_0^2\right\}e^{3\zeta/2}d\zeta+C_4,
\end{equation}
where $C_3$ and $C_4$ are arbitrary constants. We can fix $C_4$ without
loss of generality. Indeed, it is easy to see that we can add a
constant to $f(\eta)$ and the same constant to $\xi$, without changing
of Eqs. (\ref{nl4}) and (\ref{nl5}).

Following~\cite{1104.2692}, we consider the matter with the EoS
parameter $w_{\mathrm{m}}\equiv P_{\mathrm{m}}/\rho_{\mathrm{m}}$ to be a
constant, not equal to $-1$. Thus, Eq.~(\ref{equ_rho}) has the
following general solution
\begin{equation}
\rho_{\mathrm{m}}=\rho_0\,e^{{}-3(1+w_{\mathrm{m}})H_0t},
\end{equation}
where $\rho_0$ is an arbitrary constant.

Equation~(\ref{equPsi}) has the following general solution:
\begin{itemize}
\item At $\rho_0=0$,
\begin{equation}
\Psi_1(t)=C_1e^{-3H_0t}+C_2e^{-2H_0t}-1+\frac{\Lambda}{3H_0^2},
\end{equation}
\item At $\rho_0\neq 0$ and $w_{\mathrm{m}}=0$,
\begin{equation}
\Psi_2(t)=C_1e^{-3H_0t}+C_2e^{-2H_0t}-1+\frac{\Lambda}{3H_0^2}-\frac{\kappa^2\rho_0}{H_0}e^{-3H_0t}t,
\end{equation}
 \item At $\rho_0\neq 0$ and $w_{\mathrm{m}}=-1/3$,
\begin{equation}
\Psi_3(t)=C_1e^{-3H_0t}+C_2e^{-2H_0t}-1+\frac{\Lambda}{3H_0^2}+\frac{4\kappa^2\rho_0}{3H_0}e^{-2H_0t}t,
\end{equation}
\item At $\rho_0\neq 0$, $\, w_{\mathrm{m}}\neq 0$ and
$w_{\mathrm{m}}\neq -1/3$,
\begin{equation}
\Psi_4(t)=C_1e^{-3H_0t}+C_2e^{-2H_0t}-1+\frac{\Lambda}{3H_0^2}-\frac{\kappa^2\rho_0(w_{\mathrm{m}}-1)}
{3H_0^2w_{\mathrm{m}}(1+3w_{\mathrm{m}})}e^{-3H_0(w_{\mathrm{m}}+1)t},
\end{equation}
\end{itemize}
where $C_1$ and $C_2$ are arbitrary constants.

Substituting the explicit form of $\Psi(t)$, we get
\begin{itemize}

\item For the model without matter ($\rho_0=0$, $\Psi(t)=\Psi_1(t)$),
\begin{equation}
\label{f1}
f_1(\eta) = \frac{C_2}{4}e^{\eta/2}+C_3 e^{3\eta/2}+C_4,
\end{equation}
where $C_3$ and $C_4$ are arbitrary constants. Note that $C_2$ is an
arbitrary constant as well.

\item For the model with the dark matter ($w_{\mathrm{m}}=0$,
$\Psi(t)=\Psi_2(t)$),
\begin{equation}
\label{fw0} f_2(\eta)
=f_1(\eta)-\frac{\kappa^2\rho_0}{3H_0^2}e^{3\eta/4}.
\end{equation}

\item For the model, including the matter with $w_{\mathrm{m}}=-1/3$
($\Psi(t)=\Psi_3(t)$),
\begin{equation}
\label{fw13} f_3(\eta)
=f_1(\eta)+\frac{\kappa^2\rho_0}{4H_0^2}\left(1-\frac{1}{3}\eta\right)
e^{\eta/2}.
\end{equation}

\item For the model, including the matter with another value of
$w_{\mathrm{m}}$ ($\Psi(t)=\Psi_4(t)$),
\begin{equation}
\label{f4}
 f_4(\eta)
=f_1(\eta)-\frac{\kappa^2\rho_0}{3(1+3w_{\mathrm{m}})H_0^2}
e^{3(w_{\mathrm{m}}+1)\eta/4}.
\end{equation}

\end{itemize}

One can see that the key ingredient of all functions $f_i(\eta)$ is an
exponent function.

\subsection{De Sitter solutions for exponential $f(\eta)$}

In the following we consider de Sitter solutions for  the model with
\begin{equation}
\label{f} f(\eta)=f_0 e^{\eta/{\beta}}\, ,
\end{equation}
where $f_0$ and $\beta$ are constant. We choose this form of $f(\eta)$ because it is the
simplest function, which belongs to the general set of functions (\ref{f1}), (\ref{fw0}), and (\ref{f4}), described above.
Note that models involving this exponential form of $f(\eta)$ are being actively studied in the literature~\cite{Odintsov0708,Jhingan:2008ym,Koivisto:2008,Nojiri:2010pw,1104.2692,Farajollahi,ZS}.

In the case $\Lambda=0$, the corresponding system of equation can
been written in terms of Hubble-normalized variables. The stability
of the fixed point for this system has been
discussed in \cite{Odintsov0708} and the stability analysis for the same case
 has been further investigated in \cite{Jhingan:2008ym} (see
also~\cite{Farajollahi}). We will consider this specific case in
Subsection~\ref{4-2}. In
absence of matter, expanding universe solutions $a\propto t^n$  have been found in~\cite{Odintsov0708,ZS}.
In~\cite{Koivisto:2008} the ensuing cosmology at the four basic epochs:
radiation dominated, matter dominated, accelerating, and a general
scaling has been studied for an interesting nonlocal model involving, in particular, an
exponential form of $f(\eta)$. Screening of the cosmological constant in the
nonlocal model described by the action~(\ref{nl1}) with an exponential
$f(\eta)$ has been considered in~\cite{Nojiri:2010pw,1104.2692,ZS}.

In~\cite{1104.2692}, the model has been considered in further detail: the
corresponding de Sitter solutions have been obtained, a condition to avoid
a ghost has been considered, and a screening scenario for a cosmological
constant  has been discussed. Note that in~\cite{Odintsov0708,1104.2692} the authors put
restrictions on arbitrary parameters, to get de Sitter solution. It
will be interesting to get de Sitter solutions without any restriction.

Substituting the solution (\ref{eta}) of Eq.~(\ref{equ3}) into
Eq.~(\ref{equ4}) and assuming that the integration constant $\eta_0\neq
0$, we obtain that Eq.~(\ref{equ4}) has the general solution
\begin{equation}
\xi(t)
=12\frac{H_0^2f_0}{\beta}\int\limits_0^t\left[\left(C_1+\int\limits_0^{t_1}
e^{(-\eta_0\exp[{-3H_0(t_2-t_0)}]-4H_0(t_2-t_0))/\beta+3H_0t_2}dt_2
\right)e^{-3H_0t_1}dt_1\right] - \xi_0, \label{xi_sol}
\end{equation}
with $C_1$ and $\xi_0$ arbitrary constants. If $\beta=2/3$, then
$\xi(t)$ can be found explicitly:
\begin{equation}
\xi(t)=\frac{8f_0}{9\eta_0^2}e^{-(3/2)\eta_0\exp(-3H_0(t-t_0))}-C_1e^{-3H_0(t-t_0)}-\xi_0.
\end{equation}
The solutions obtained include four arbitrary parameters, namely
$\eta_0$, $\xi_0$, $C_1$, and $t_0$. As we have mentioned above one can set
$t_0=0$, without loss of generality.

At $H(t)=H_0$, Eq.~(\ref{equ12}) has the  form
\begin{equation}
 \dot\xi \dot\eta + \frac{1}{\beta}f(\eta)\left(\frac{1}{\beta}\dot\eta^2+\ddot\eta\right)
 - \frac{1}{\beta}H_0f(\eta)\dot\eta -
 \ddot\xi+H_0\dot\xi + \kappa^2 (1+w_\mathrm{m})\rho_{\mathrm{m}}=0\, ,
\end{equation}
and, using (\ref{equ4}),
\begin{equation}\label{equ12H0}
( \dot\eta+4H_0) \dot\xi +
\frac{f(\eta)}{\beta}\left(\frac{1}{\beta}\dot\eta^2+\ddot\eta
 - H_0\dot\eta - 12H_0^2\right) + \kappa^2 (1+w_\mathrm{m})\rho_{\mathrm{m}}=0\, .
\end{equation}
Substituting the explicit
expression for $\eta$, we get
\begin{equation*}
 3H_0\eta_0e^{-3H_0t}\dot\xi + \frac{H_0^2}{\beta^2}f(\eta)
 \left(9\eta_0^2e^{-6H_0t}-12\eta_0(\beta+2)e^{-3H_0t}-8(\beta-2)\right)
 + \kappa^2 (1+w_\mathrm{m})\rho_{\mathrm{m}}=0\, ,
\end{equation*}
where
\begin{equation*}
f(\eta)=f_0e^{-4H_0t/\beta}e^{-\eta_0\exp({-3H_0t})/\beta},\qquad
\rho_{\mathrm{m}}=\rho_0\,e^{{}-3(1+w_{\mathrm{m}})H_0 t}.
\end{equation*}
From (\ref{xi_sol}), it follows that
\begin{equation}
\dot\xi=12\frac{H_0^2f_0}{\beta}e^{-3H_0t}\left(C_1+\int\limits_0^{t}
e^{[-\eta_0\exp(-3H_0t_2)-4H_0t_2]/\beta+3H_0t_2}\, dt_2
\right).
\end{equation}
A straightforward calculation shows that Eq.~(\ref{equ12H0})
has no solution for any value of the parameters such that $f_0\neq 0$,
$\eta_0\neq 0$, and $H_0\neq 0$.

Therefore, without loss of generality we can put $\eta_0=0$. In this
case, for $\beta\neq 4/3$, from (\ref{equ3}) and (\ref{equ4}) the
following solution is obtained~\cite{1104.2692}:
\begin{eqnarray}\label{solxi_eta}
    \xi={}-\frac{3f_0\beta}{3\beta-4}e^{-4H_0(t-t_0)/\beta}+\frac{c_0}{3H_0}e^{-3H_0(t-t_0)}-\xi_0,\qquad
    \eta={}-4H_0(t-t_0),
\end{eqnarray}
where $c_0$ is an arbitrary constant,
\begin{equation}
\Lambda = 3H_0^2(1+\xi_0),\qquad \rho_0 = \frac{6
\left(\beta-2\right) H_0^2f_0}{\kappa^2\beta}\ ,\qquad w_{\mathrm{m}} =
{}-1+\frac{4}{3\beta}.
\label{CondeSit}
\end{equation}
The case $\beta=2$ corresponds to $\rho_0=0$. Note that $\beta=2$
corresponds to $w_{\mathrm{m}} =-1/3$. It means that the model with
exponential $f(\eta)$ has no de Sitter solutions if we add this kind
of matter. The type of function $f(\eta)$ that can have such solutions
is given by (\ref{fw13}).

For $\beta=4/3$, we get
\begin{equation}
\xi(t) = {}-f_0(c_0+3H_0(t-t_0))e^{-3H_0(t-t_0)}- \xi_0,
\end{equation}
\begin{equation}
\Lambda = 3H_0^2(1+\xi_0),\qquad P_{\mathrm{m}}=0, \qquad
\rho_{\mathrm{m}}={}-\frac{3}{\kappa^2}H_0^2f_0e^{-3H_0(t-t_0)}.
\end{equation}
This solution clearly corresponds to dark matter, because
$w_{\mathrm{m}}=0$.

\section{Stability of the de Sitter background}

\subsection{The case of nonzero $\Lambda$}
\subsubsection{The FLRW metric}

Let us now introduce new variables
\begin{equation}
\phi=f(\eta)=f_0 e^{\eta/\beta}, \qquad \psi=\dot\eta.
\end{equation}
The functions $\phi(t)$ and $\psi(t)$ are connected by the
equation
\begin{equation}
\label{equphi}
\dot\phi=\frac{1}{\beta}\phi\psi.
\end{equation}
The system (\ref{equ12})--(\ref{equ4}) can be expressed as a
system of first-order differential equations, in terms of new
variables. We rewrite Eqs.~(\ref{equ3}) and (\ref{equ4}) as
\begin{equation}
\label{equ3n} \dot \psi={}-  3H\psi- 6\left(\dot H + 2
H^2\right) \, ,
\end{equation}
\begin{equation}
\label{equ4n}\dot \xi= \vartheta, \qquad  \dot \vartheta={} - 3H
\vartheta+\frac{6}{\beta}\left( \dot H + 2 H^2\right)\phi \,.
\end{equation}
Using
\begin{equation*}
\ddot\phi=\frac{\phi}{\beta}\left(\frac{\psi^2}{\beta}-3H\psi-6\dot H-12H^2\right),
\end{equation*}
we get that Eq.~(\ref{equ12}) is equivalent to
\begin{equation}
\label{equ12n} 2\dot H \left(1 + \frac{\beta-6}{\beta}\phi -
\xi\right)=4H\left(\frac{\phi\psi}{\beta}-\vartheta\right){}-\frac{1}{\beta^2}\phi\psi^2
+\frac{24}{\beta}H^2\phi - \vartheta \psi - \kappa^2
(1+w_{\mathrm{m}})\rho_{\mathrm{m}}.
\end{equation}

Consider the de Sitter solution
\begin{eqnarray}\label{rhp,P}
    &&\rho_{\mathrm{m}}=\rho_0e^{-3(w_{\mathrm{m}}+1)H_0(t-t_0)},\quad
    P_{\mathrm{m}}=w_{\mathrm{m}}\rho_{\mathrm{m}},\quad\, \Lambda=3H_0^2(1+\xi_0), \\&&
    \beta=\frac{4}{3(1+w_{\mathrm{m}})},\quad
    \psi={}-4H_0,\quad\phi=f_0 e^{-4H_0t/\beta}.
\end{eqnarray}
For $\beta\neq 4/3$, we have
\begin{equation*}
\xi={}-\frac{3f_0\beta}{3\beta-4}e^{-4H_0(t-t_0)/\beta}+\frac{c_0}{3H_0}e^{-3H_0(t-t_0)}-\xi_0,
\end{equation*}
and, for $\beta=4/3$,
\begin{equation*}
\xi={}-f_0(c_0+3H_0(t-t_0))e^{-3H_0(t-t_0)}- \xi_0.
\end{equation*}
As $t$ tends to $+\infty$,
\begin{equation}
\rho_{\mathrm{m}} \rightarrow 0,\qquad \phi\rightarrow 0,\qquad\psi= {}-4H_0,
\qquad\xi\rightarrow {}-\xi_0,
\end{equation}
for all $H_0>0$ and $\beta>0$. This system  has a fixed point:
\begin{equation*}
\phi=0,\qquad \xi={}-\xi_0,\qquad \psi={}-4H_0, \qquad \rho_{\mathrm{m}}=0.
\end{equation*}
 Note that  we cannot
fix $H_0$, using the relation $\Lambda=3H_0^2(1+\xi_0)$,  since $\xi_0$
is an arbitrary parameter. So, we have no isolated fixed point.

Two different cases appear: $\Lambda=0$ and $\Lambda\neq 0$. In this subsection, we consider
the case $\Lambda\neq 0$. For $\Lambda=0$, the stability can be analyzed using a change of variables.
This analysis will be presented in the next subsection.

 For $\Lambda\neq
0$, one gets $\xi_0\neq -1$, in the neighborhood of the fixed point
\begin{equation}
\label{Condi}
\left(1 + \frac{\beta-6}{\beta}\phi - \xi\right)\approx 1+\xi_0\neq 0
\end{equation}
and we can divide Eq.~(\ref{equ12n}) in this expression to get the
equation in the standard form:
\begin{equation}
\label{equ12n2}
 \dot H =\frac{1}{2\left(1 + \frac{\beta-6}{\beta}\phi-
\xi\right)}\left[4H\left(\frac{\phi\psi}{\beta}-\vartheta\right)-\frac{\phi\psi^2}{\beta^2}
+\frac{24}{\beta}H^2\phi - \vartheta \psi - \frac{4\kappa^2}{3\beta}
\rho_{\mathrm{m}}\right].
\end{equation}
We consider the time domain $(t_1,+\infty)$ such that  $1 + \frac{\beta-6}{\beta}\phi - \xi\neq 0$.
This expression is positive for $\Lambda>0$ and negative for $\Lambda<0$.
In the neighborhood of the fixed point, which corresponds to de Sitter solution, we have
\begin{eqnarray}
H(t)&=&H_0 + \varepsilon h_1(t) +{\cal O}(\varepsilon^2), \label{m}
\\
\phi(t)&=&\varepsilon  \phi_1(t)+{\cal O}(\varepsilon^2), \label{n}
\\
\psi(t)&=&{}-4H_0+\varepsilon \psi_1(t)+{\cal O}(\varepsilon^2),
\label{o}
\\
\xi(t)&=&{}-\xi_{0}+\varepsilon \xi_1(t)+{\cal O}(\varepsilon^2),
\label{p}
\\
\vartheta(t)&=&\varepsilon \vartheta_1(t)+{\cal O}(\varepsilon^2)
\label{q},\\
\rho_{\mathrm{m}}(t)&=&\varepsilon \rho_{m1}(t)+{\cal O}(\varepsilon^2)
\label{rhom},
\end{eqnarray}
where $\varepsilon$ is a small parameter.

From Eqs.~(\ref{equ_rho}), (\ref{equphi}), (\ref{equ3n}),
(\ref{equ4n}),  and (\ref{equ12n2}), we obtain, to first order in
$\varepsilon$,  the following system:
\begin{eqnarray}
\displaystyle \dot\rho_{m1}&=&\displaystyle
{}-\frac{4}{\beta}H_0\rho_{m1}, \label{u}
\\
\displaystyle \dot\phi_1&=&\displaystyle {}-\frac{4}{\beta}H_0\phi_1,
\label{s}
\\
\displaystyle \dot\vartheta_1&=&\displaystyle{}
-3H_0\vartheta_1+\frac{12}{\beta}H_0^2\phi_1, \label{v}
\\
\displaystyle \dot h_1&=&\displaystyle
{}\frac{2}{(1+\xi_0)}\left[\frac{2}{\beta}\left(1-\frac{2}{\beta}\right)H_0^2\phi_1-
\frac{\kappa^2}{3\beta}\rho_{m1}\right],
\label{h1}
\\
 \displaystyle
\dot\psi_1&=&\displaystyle{}-3H_0\psi_1-12H_0h_1-
{}\frac{12}{(1+\xi_0)}\left[\frac{2}{\beta}\left(1-\frac{2}{\beta}\right)H_0^2\phi_1-
\frac{\kappa^2}{3\beta}\rho_{m1}\right]. \label{t}
\end{eqnarray}
Note that the function $\xi_1$ is not included in this system. It can
be defined using Eq.~(\ref{equ1}). It is plain that $\xi_1$ cannot
tend to infinity, if all other first-order corrections are bounded.

Let us now consider the system (\ref{u})--(\ref{t}).
The functions
\begin{equation}
\label{rho1phi1}
\rho_{m1}(t)=d_0 e^{-4H_0t/{\beta}}, \quad \phi_1(t)=d_1e^{-4H_0t/{\beta}},
\end{equation}
where $d_0$ and $d_1$ are arbitrary constants, are general solutions of
Eqs. (\ref{u}) and (\ref{s}), respectively. At $H/\beta>0$, these
functions tend to zero, for $t\rightarrow\infty$. Substituting these
functions into the other equations, we get
\begin{equation}
\label{vartheta1}
\vartheta_1(t)=12\frac{H_0d_1}{3\beta-4}e^{-4H_0t/\beta}+d_3e^{-3H_0t},
\end{equation}
\begin{equation}
h_1(t)=d_2-\frac{6H_0^2d_1(\beta-2)-\kappa^2d_0\beta}{6\beta H_0(1+\xi_0)}e^{-4H_0t/\beta},
\end{equation}
\begin{equation}
\psi_1(t)=\frac{2(\beta-2)(6H_0^2\beta d_1-12H_0^2d_1-\kappa^2\beta d_0)}
{H_0\beta(3\beta-4)(1+\xi_0)}e^{-4H_0t/\beta}+ d_4e^{-3H_0t}-4d_2,
\end{equation}
where $d_2$, $d_3$, and $d_4$ are arbitrary constants.
The two last expressions are valid for $\beta\neq 4/3$.
For $\beta= 4/3$,
 \begin{equation*}
\vartheta_1=\left(9H_0^2d_1t+d_3\right)e^{-3H_0t},
\quad\psi_1=\left(\frac{(3H_0^2d_1+\kappa^2d_0)t}{1+\xi_0}+ d_4\right)e^{-3H_0t}-4d_2.
\end{equation*}
We see that none of the perturbations tends to infinity at
$t\rightarrow\infty$, so that the de Sitter solutions are stable. We
should note that the fixed point, which corresponds to a de Sitter
solution with fixed $H_0$, is not isolated, because there is only one
condition: $\Lambda = 3H_0^2(1+\xi_0)$ on the two arbitrary parameters
$H_0$ and $\xi_0$. Changing $\xi_0$, we get a new value of  $H_0$ for
the same $\Lambda$. This is the reason why the function $h_1(t)$ has an
arbitrary parameter $d_2$. To analyze the stability of such a fixed
point one cannot use Lyapunov's theorem~\cite{LyapunovPontryagin}.

\subsubsection{The Bianchi I metric}

The Bianchi universe models are spatially
homogeneous anisotropic cosmological models. There are strong limits on
anisotropic models from observations~\cite{Bernui:2005pz}.
Anisotropic spatially homogeneous fluctuations have to be strongly
suppressed, and models developing large anisotropies should be discarded
as early or late times cosmological models.

Interpreting the solutions of the Friedmann equations as isotropic
solutions in the Bianchi I metric, we include anisotropic perturbations
in our consideration. A similar stability analysis has been made for
cosmological models with scalar fields and phantom scalar fields
in~\cite{ABJV0903}.  The stability analysis is essentially simplified
by a suitable choice of variables. Let us consider the Bianchi I metric
\begin{equation} \label{Bianchi}
{ds}^{2}={}-{dt}^2+a_1^2(t)dx_1^2+a_2^2(t)dx_2^2+a_3^2(t)dx_3^2.
\end{equation}
It is convenient to express $a_i$ in terms of new variables $a$ and
$\beta_i$ (we use the notation of~\cite{Pereira}):
\begin{equation}
a_i(t)= a(t) e^{\beta_i(t)}.
\end{equation}
Imposing the constraint
\begin{equation}
\label{restr1} \beta_1(t)+\beta_2(t)+\beta_3(t)=0,
\end{equation}
at any $t$, one has the following relations
\begin{equation}
a(t)=(a_1(t)a_2(t)a_3(t))^{1/3},
\end{equation}
\begin{equation}\label{Hi}
H_i\equiv \frac{\dot a_i}{a_i}= H+\dot\beta_i, \qquad\mbox{and}\qquad H\equiv
\frac{\dot a}{a}=\frac{1}{3}(H_1+H_2+H_3).
\end{equation}
Note that $\beta_i$ are not components of a vector and, therefore,
are not subject to the Einstein summation rule. In the case of
the FLRW metric: $a_1=a_2=a_3=a$, all $\beta_i$ are equal to zero and $H$ is the
Hubble parameter;  thus,
we can use the same notations as for the FLRW metric.  Following~\cite{Pereira},  we introduce the shear
\begin{equation}
\sigma^2\equiv \dot\beta_1^2+\dot\beta_2^2+\dot\beta_3^2.
\end{equation}
It is easy to calculate that
\begin{equation}
R=12H^2+6\dot H+\sigma^2,
\end{equation}
\begin{equation}
R_{00}=-3\left(H^2+\dot H\right)-\sigma^2,
\end{equation}
\begin{equation}
R_{jj}=g_{jj}(\dot H_j+3H H_j)=g_{jj}\left(\dot H+3H^2+\ddot\beta_j+3H\dot\beta_j\right).
\end{equation}

In the Bianchi I metric, we get $\Box\eta=R$ as
\begin{equation}
\label{equpsiB}
\dot\psi={}-3H\psi-12H^2-6\dot H-\sigma^2.
\end{equation}
Equations~(\ref{nl5}) and (\ref{equ_rho}) are now
\begin{equation}
\label{equthetaB}
    \dot \vartheta={}-3H \vartheta+\frac{\phi}{\beta}\left(12H^2+6\dot H+\sigma^2\right),
\end{equation}
\begin{equation}
\label{equ_rhoB} \dot\rho_{\mathrm{m}}={}-
\frac{4}{\beta}H\rho_{\mathrm{m}}.
\end{equation}
To get Eq.~(\ref{equ_rhoB}) we have used  condition (\ref{CondeSit}) on $w_\mathrm{m}$.
The Einstein equations have the form:
\begin{equation}
\label{equ1B}
{}\left[\frac{\sigma^2}{2}- 3 H^2\right]\!\left(1 + \phi - \xi\right) + \frac{1}{2}\dot\xi \psi
 - 3H\left(\dot\phi -\dot\xi \right)  + \Lambda
+ \kappa^2 \rho_{\mathrm{m}}=0\, ,
\end{equation}
\begin{equation}
\label{equ2B}
\left[2\dot H + 3H^2+\frac{\sigma^2}{2}-\ddot\beta_j-3H\dot\beta_j\right]\! \left(1 + \phi - \xi\right)
+ \frac{1}{2}\dot\xi\psi
+ \ddot\phi-\ddot\xi + (2H-\dot\beta_j)(\dot\phi -\dot\xi)  = \Lambda - \kappa^2 P_{\mathrm{m}}.
\end{equation}
Adding Eqs.~(\ref{equ2B}) for $j=1,2,3$ and using (\ref{restr1}), we get
\begin{equation}
\label{equ2Bsum}
\left[2\dot H + 3H^2+\frac{\sigma^2}{2}\right]\! \left(1 + \phi - \xi\right)
+ \frac{1}{2}\dot\xi \psi
+ \ddot\phi-\ddot\xi + 2H\left(\dot\phi -\dot\xi\right)  = \Lambda - \kappa^2 P_{\mathrm{m}}
\end{equation}
and adding now (\ref{equ1B}) and (\ref{equ2Bsum}),
\begin{equation}
\label{equdotHB}
\left[2\dot H +\sigma^2\right]\! \left(1 + \phi - \xi\right)
+ \dot\xi \psi
+ \ddot\phi-\ddot\xi - H(\dot\phi -\dot\xi)  = - \kappa^2(1+w_{\mathrm{m}}) \rho_{\mathrm{m}}.
\end{equation}
Assuming that (\ref{Condi}) is satisfied, we write Eq.~(\ref{equdotHB}) as
\begin{equation}
\label{equ12n2B}
\begin{split}
 \dot H &=\frac{1}{2\left[1 + \frac{\beta-6}{\beta}\phi-
\xi\right]}\left(4H\left[\frac{\phi\psi}{\beta}-\vartheta\right]-\frac{\phi\psi^2}{\beta^2}+{}\right.\\
&\left.{}+\frac{24}{\beta}H^2\phi - \vartheta \psi - \frac{4\kappa^2}{3\beta}
\rho_{\mathrm{m}}-\left[1 + \frac{\beta-2}{\beta}\phi-
\xi\right]\sigma^2\right).
\end{split}
\end{equation}
Subtracting (\ref{equ2B}), with $j=i$, from (\ref{equ2B}), with $j=k$, we obtain the following system of equations:
\begin{equation}
\label{SYSTEMbeta}
\left[\ddot\beta_i+3H\dot\beta_i-\ddot\beta_k-3H\beta_k\right]\! \left(1 + \phi - \xi\right)+(\dot\beta_i-\dot\beta_k)
(\dot\phi -\dot\xi)=0
\end{equation}
and using (\ref{restr1}), it is easy to get from this system
\begin{equation}
\left[\ddot\beta_i+3H\dot\beta_i\right]\! \left(1 + \phi -
\xi\right)+\dot\beta_i\left(\frac{1}{\beta}\phi\psi -\vartheta\right)=0
\label{equbeta}
\end{equation}
and
\begin{equation}
\label{equvartheta}
\left[\frac{d}{dt}\left(\sigma^2\right)+6H\sigma^2\right]\! \left(1 +
\phi - \xi\right)+2\sigma^2\left(\frac{1}{\beta}\phi\psi
-\vartheta\right)=0.
\end{equation}
The functions $H(t)$, $\sigma^2(t)$, $\phi(t)$, $\psi(t)$, $\xi(t)$, $\vartheta(t)$ and $\rho_{\mathrm{m}}(t)$ can
be obtained from equations (\ref{equphi}), (\ref{equpsiB})--(\ref{equ_rhoB}), (\ref{equ12n2B})  and
(\ref{equvartheta}). If $H(t)$ and the scalar fields are known, then $\beta_i$ can be directly
obtained from (\ref{equbeta}).

The functions $H(t)$, $\dot\beta_i(t)$, and $\sigma^2(t)$ are very
suitable to analyze the stability of isotropic solutions in the Bianchi
I metric. Indeed, the use of these variables makes the analysis of
stability in the FLRW and Bianchi I metrics similar, because the
equations of motion  in the Bianchi I metric with $\sigma^2=0$ are
identical to the equations of motion  in the FLRW metric. Thus, we can
use some results from the previous subsection. In the neighborhood of
the fixed point, which corresponds to de Sitter solution, we have the
expansions (\ref{m})--(\ref{rhom}), and
\begin{equation}\label{sigma}
    \sigma^2(t)=\varepsilon \sigma^2_1(t)+{\cal O}(\varepsilon^2).
\end{equation}
To first order in $\varepsilon$, we get the following system of
equations. Equations (\ref{u})--(\ref{v}) are valid for the Bianchi I
metric as well, and instead of Eqs. (\ref{h1}) and (\ref{t}) we get the
system:
\begin{eqnarray}
\displaystyle \dot h_1&=&\displaystyle
{}\frac{2}{(1+\xi_0)}\left[\frac{2}{\beta}\left(1-\frac{2}{\beta}\right)H_0^2\phi_1-
\frac{\kappa^2}{3\beta}\rho_{m1}\right]-\frac12\sigma^2_1,
\label{h1B}
\\
 \displaystyle
\dot\psi_1&=&\displaystyle{}-3H_0\psi_1-12H_0h_1+2\sigma^2_1-
{}\frac{12}{(1+\xi_0)}\left[\frac{2}{\beta}\left(1-\frac{2}{\beta}\right)H_0^2\phi_1-
\frac{\kappa^2}{3\beta}\rho_{m1}\right], \label{tB}\\
\displaystyle
\frac{d}{dt}(\sigma^2_1)&=&\displaystyle{}-6H_0\sigma^2_1.\label{sigma2_1}
\end{eqnarray}
From the last equation, we get
\begin{equation}
\sigma^2_1=d_5e^{-6H_0t},
\end{equation}
where $d_5$ is an arbitrary constant.

The expressions for the functions $\phi_1(t)$, $\rho_1(t)$ and
$\vartheta_1(t)$ in the Bianchi I metric coincide with the
corresponding expressions in the FLRW metric, which are given by
(\ref{rho1phi1})--(\ref{vartheta1}). Substituting these functions into
Eqs.~(\ref{h1B}) and (\ref{tB}), we obtain
\begin{equation}
    h_1=d_2-\frac{6H_0^2d_1(\beta-2)-\kappa^2d_0\beta}{6\beta H_0(1+\xi_0)}e^{-4H_0t/\beta}+\frac{d_5}{12H_0}e^{-6H_0t},
\end{equation}
\begin{equation}
    \psi_1=\frac{2(\beta-2)(6H_0^2\beta d_1-12H_0^2d_1-\kappa^2\beta d_0)}
{H_0\beta(3\beta-4)(1+\xi_0)}e^{-4H_0t/\beta}+ d_4e^{-3H_0t}-4d_2-\frac{d_5}{3H_0}e^{-6H_0t}
\end{equation}
and, at $\beta=4/3$, we get
\begin{equation*}
\psi_1=\left(\frac{\left(3H_0^2d_1+\kappa^2d_0\right)t}{1+\xi_0}+ d_4\right)e^{-3H_0t}-4d_2-\frac{d_5}{3H_0}e^{-6H_0t}.
\end{equation*}

The function $\xi_1(t)$, which can be defined using Eq.~(\ref{equ1B}),
is a bounded function if all other first-order corrections are bounded.
Thus, we come to the conclusion that de Sitter solutions are stable if
$H_0>0$ and $\beta>0$. So, the stability conditions in the cases of the
FLRW and Bianchi I metrics coincide.

\subsection{The case $\Lambda=0$. Normalized variables}
\label{4-2}

To analyze the stability of the de Sitter solutions at $\Lambda=0$,
following~\cite{Odintsov0708} we transform the system of equations
using new dependent variables
\begin{equation}
X={}-\frac{\dot{\eta}}{4H},\qquad
  W= \frac{\dot{\xi}}{6Hf},\qquad
  Y=\frac{1-\xi}{3f},\qquad
  Z=\frac{\kappa^2\rho_{\mathrm{m}}}{3H^2f}
\end{equation}
and the independent variable $N$:
\begin{equation}
\frac{d}{dN}\equiv a\frac{d}{da}= \frac{1}{H}\frac{d}{dt}\ .
\end{equation}
The use of the Hubble-normalized variables~\cite{Wainwright_Lim} and
$N$ as  time variable makes the equation of motion dimensionless. Note
that a change of dependent and independent variables of this kind is
actively used in cosmological models with scalar fields, in order to
analyze the stability in the FLRW
metric~\cite{phantom-attractor,Lazkoz}, as well as in models of
inflation (see~\cite{Tegmark} and references therein). Clearly,
  \begin{eqnarray}
    \frac{d X}{d N}&=&\frac{1}{H}
    \frac{d X}{d t}={}-\frac{1}{4H}\left(\frac{\ddot{\eta}}{H}-\frac{\dot{H}}{H^2}\dot{\eta}\right)=
    {}-\frac{\ddot\eta}{4H^2}-\frac{X}{H}\frac{dH}{dN},\label{dX}\\
   \frac{dW}{dN}&=&\frac{1}{H}\left(\frac{\ddot{\xi}}{6Hf}-\frac{\dot{\xi}\dot{H}}{6H^2f}
-\frac{\dot{\xi}\dot{f}}{6Hf^2}\right)=\frac{\ddot\xi}{6fH^2}+\frac{4}{\beta}XW-\frac{W}{H}\frac{dH}{dN}.\label{dW}
 \end{eqnarray}
Equations~(\ref{equ3}) and (\ref{equ4}) are equivalent to the following
ones, in terms of the new variables,
\begin{eqnarray}
\frac{dX}{dN}&=&3(1-X)+\frac{1}{H}\left(\frac{3}{2}-X\right)\frac{dH}{dN}\,,\label{dXsys} \\
\frac{dW}{dN}&=&\frac{2}{\beta}(1+2WX)-3W+\frac{1}{H}
\left(\frac{1}{\beta}-W\right)\frac{dH}{dN}\,\label{dWsys}
\end{eqnarray}
and Eq.~(\ref{equ_rho}) can be written as
\begin{equation}
\frac{dZ}{dN}=\frac{4}{\beta}(X-1)Z-2\frac{Z}{H}\frac{dH}{dN}\,.\label{dZsys}\\
\end{equation}
To get the full system of equations we need one for
$\frac{dH}{dN}$. In terms of the new variables, Eq.~(\ref{equ1}) has
the form
\begin{equation}
6W-3Y-4XW+\frac{4}{\beta}X+Z=1. \label{equ1nv}
\end{equation}
 Differentiating it 
\begin{eqnarray}
6\frac{dW}{dN}-3\frac{dY}{dN}-4W\frac{dX}{dN}-4X\frac{dW}{dN}+\frac{4}{\beta}\frac{dX}{dN}
+\frac{dZ}{dN}=0,\label{dEqInit}
\end{eqnarray}
using
\begin{equation*}
\frac{dY}{dN}=2\left(\frac{2XY}{\beta}-{W}\right),\label{dY}
\end{equation*}
and substituting \eqref{dXsys}--\eqref{equ1nv},  we get
\begin{equation}
\left(4WX-6W-\frac{4}{\beta}X+\frac{6}{\beta}-Z\right)\frac{1}{H}\frac{dH}{dN}
+12W(X-1)+\frac{2}{\beta}(6-4X-Z)-\frac{8}{\beta^2}X^2=0.
\label{dEquation}
\end{equation}
For $c_0=0$ (and $\beta\neq 4/3$) the de Sitter solution in terms of new variables
corresponds to the following fixed point:
\begin{equation}\label{FixPoi}
H=H_0,\qquad W_0=\frac{2}{3\beta-4},\qquad X_0=1,\qquad Y_0=\frac{\beta}{3\beta-4},\qquad
Z_0=\frac{2(\beta-2)}{\beta}\,.
\end{equation}
For $\beta=2$, when $\rho_0=0$, the
stability of the de Sitter solution has been proven in~\cite{Odintsov0708}.
In this paper, we discuss stability in the case  $\beta>0$.

Let us consider perturbations in the neighborhood of the de Sitter
solution (\ref{FixPoi}):
\begin{equation*}
X=X_0(1+\varepsilon x_1(N)) +{\cal O}(\varepsilon^2),\qquad Z=Z_0(1+\varepsilon z_1(N)) +{\cal O}(\varepsilon^2),
\end{equation*}
\begin{equation*}
W=W_0(1+\varepsilon w_1(N)) +{\cal O}(\varepsilon^2),\qquad H=H_0(1+\varepsilon h_1(N)) +{\cal O}(\varepsilon^2).
\end{equation*}
To first order in $\varepsilon$, we obtain the system of
linear equations:
\begin{equation}
\label{equx1} \frac{dx_1}{dN}={}-3x_1+\frac{1}{2}\frac{dh_1}{dN},
\end{equation}
\begin{equation}
\label{equz1} \frac{dz_1}{dN}=\frac{4}{\beta}x_1-2\frac{dh_1}{dN},
\end{equation}
\begin{equation}
\label{equw1}
\frac{dw_1}{dN}=\frac{4}{\beta}x_1-\frac{4-\beta}{2\beta}\frac{dh_1}{dN}+\left(\frac{4}{\beta}-3\right)w_1,
\end{equation}
\begin{equation}
\label{equh1}
\frac{dh_1}{dN}={}-\frac{8(\beta-4)}{\beta(3\beta^2-11\beta+12)}x_1
-\frac{2(3\beta-4)(\beta-2)}{\beta(3\beta^2-11\beta+12)}z_1.
\end{equation}
Substituting (\ref{equh1}) into (\ref{equx1}) and (\ref{equz1}), we get
\begin{equation}
\label{systx1z1}
\begin{split}
\frac{dx_1}{dN}&={}-\frac{(\beta-1)(3\beta-4)^2}{\beta(3\beta^2-11\beta+12)}x_1
-\frac{(3\beta-4)(\beta-2)}{\beta(3\beta^2-11\beta+12)}z_1, \\
\frac{dz_1}{dN}&=\frac{4(3\beta^2-7\beta-4)}{\beta(3\beta^2-11\beta+12)}x_1+
\frac{4(3\beta-4)(\beta-2)}{\beta(3\beta^2-11\beta+12)}z_1,
\end{split}
\end{equation}
which has the following solution:
\begin{equation}
x_1=c_1e^{\lambda_1 N}+c_2e^{\lambda_2 N}, \qquad z_1=c_3e^{\lambda_1
N}+c_4e^{\lambda_2 N},
\end{equation}
where $c_1$ and $c_2$ are arbitrary constants
\begin{equation*}
c_3={}-\frac{9\beta^3-21\beta^2+16-D}{2(3\beta-4)(\beta-2)}c_1,\qquad
c_4={}-\frac{9\beta^3-21\beta^2+16+D}{2(3\beta-4)(\beta-2)}c_2,
\end{equation*}
\begin{equation*}
\lambda_1={}-\frac{9\beta^3-45\beta^2+80\beta-48+D}{2\beta(3\beta^2-11\beta+12)},\qquad
\lambda_2={}-\frac{9\beta^3-45\beta^2+80\beta-48-D}{2\beta(3\beta^2-11\beta+12)},
\end{equation*}
\begin{equation*}
D=\sqrt{81\beta^6-378\beta^5+297\beta^4+1104\beta^3-1984\beta^2+256\beta
+768}.
\end{equation*}
We see that $\lambda_1=0$ at
\begin{equation}
\beta_1=\frac{4}{3},\quad \beta_2=2,\quad
\beta_{3,4}=\frac{11}{6}\pm\frac{\sqrt{23}}{6}i.
\end{equation}
The real part of $\lambda_1$ is negative in the interval
$\beta\in(4/3,2)$. It is easy to show that also the real part of
$\lambda_2$ is negative in this interval.

Therefore, the perturbations $x_1$ and $z_1$ decrease in $4/3<\beta<2$.
Substituting $x_1(N)$ and $z_1(N)$ into Eqs.~(\ref{equw1}) and
(\ref{equh1}), we get that $h_1(N)$ and $w_1(N)$ decrease as well. Note
that $h_1(N)$ has a part $H_{1}$, which does not depend on $N$, and
therefore can be considered as part of $H_0$. This result corresponds
to the fact that, for $\Lambda=0$, the value of $H_0$ can be selected
arbitrarily; thus, one can choose $\tilde{H}_0=H_0+H_{1}$ instead of
$H_0$. Adding here the results of \cite{Odintsov0708}, we can summarize
that the de Sitter solutions are stable with respect to perturbations
in the FLRW metric with $4/3<\beta\leqslant 2$. At $f_0>0$ the stable de Sitter
solution corresponds to $\rho_0\leqslant 0$.

Consider now the case of an arbitrary $c_0$. For the de Sitter solution, we get
\begin{equation*}
H=H_0,\qquad X_0=1,\qquad
Z_0=\frac{2(\beta-2)}{\beta}\,,
\end{equation*}
\begin{equation*}
Y=\frac{\beta}{3\beta-4}-\frac{c_0}{9H_0f_0}e^{-(3-4/\beta)(N-N_0)},
\qquad W=\frac{2}{3\beta-4}-\frac{c_0}{6H_0f_0}e^{-(3-4/\beta)(N-N_0)},
\end{equation*}
where $N_0=H_0t_0$.
For $\beta>4/3$,
\begin{equation*}
\lim_{N\rightarrow\infty} Y=\frac{\beta}{3\beta-4}, \qquad \lim_{N\rightarrow\infty}  W=\frac{2}{3\beta-4}.
\end{equation*}
Thus, the de Sitter solution tends to a fixed point, which means that,
for any $\varepsilon>0$, there exists a such number $N_1$ that the de Sitter
solution is in the $\varepsilon/2$ neighborhood of the fixed point for all $N>N_1$.
Therefore, the stability of the fixed point guarantees the stability of
de Sitter solutions for any value of $c_0$. We reach the conclusion
that, at $\Lambda=0$, all de Sitter solutions are stable for
$4/3<\beta\leqslant 2$.

For $\beta=4/3$, we have no fixed point, because the $Y$ and $W$
corresponding to de Sitter solutions are proportional to $N$. Thus,
this choice of dimensionless variable is not suitable to consider
stability of the de Sitter solutions for  $\beta=4/3$. For $\beta<4/3$
and $\beta>2$, we find that the solutions are in fact unstable.

\section{Conclusions}

In this paper, we have investigated de Sitter solutions in a nonlocal
gravity model, which is described by
the action given in (\ref{nl1}) (see also \cite{Odintsov0708}).
To carry out this task we have used
the local formulation of the model (\ref{anl2}), which includes two
scalar fields. De Sitter solutions play a very important role in
cosmological models, because both inflation and the late-time Universe
acceleration can be described as a de Sitter solution with
perturbations.

We have found the interesting result that the model has de Sitter solutions
only if $f(\eta)$ satisfies the second-order linear differential
equation (\ref{equaf}). If we consider models without matter or with a
perfect matter fluid with a constant EoS parameter
$w_{\mathrm{m}}\neq -1/3$, then $f(\eta)$ can be an exponential function
or a sum of exponential functions. For the model with $f(\eta)$ equal to a
sum of exponential functions, particular de Sitter solutions
have been found in~\cite{Odintsov0708,1104.2692}.

In this paper, we have considered de Sitter solutions for an exponential
$f(\eta)$ and found all solutions in the FLRW metric that
correspond to a constant, nonzero value of the Hubble parameter,
$H_0$. In particular, we have obtained the de Sitter solutions in the
case when the matter included in the model is dark matter. This case
has never been considered before in the literature.

When $t\rightarrow \infty$, the de Sitter solutions tend to fixed
points. In the model considered, the value of the cosmological constant
does not fix $H_0$, therefore, the fixed points that correspond to de
Sitter solutions are not isolated. We have analyzed the stability of
these solutions in the cases of FLRW and Bianchi I metrics and obtained
that the first-order corrections have no increasing modes\footnote{In the case $\Lambda=0$, we have analyzed stability
in the FLRW metric only, in other words, with respect to isotropic perturbations. In our new paper~\cite{EPVQFTHEP2011}, we have shown that de Sitter solutions, which are stable with respect to isotropic perturbations, are  stable with respect to anisotropic perturbations of the Bianchi~I metric as well.}, this being
valid for any nonzero value of $\Lambda$, and for $H_0>0$ and $\beta>0$.
They display constant and decreasing modes. For this reason, we can say
that, for  $H_0>0$ and $\beta>0$, our solutions are stable for all
nonzero values of $\Lambda$. For $\Lambda=0$, the stable solutions
correspond to $H_0>0$ and $4/3<\beta\leqslant 2$.

Looking further, it will be interesting to
consider the stability of the de Sitter solutions and the corresponding
ghost-free conditions in the Einstein frame, for models with more general
functions $f(\eta)$ satisfying the differential equation (\ref{equaf}).
\medskip

\noindent {\bf Acknowledgements.} The authors thank Sergei~D.~Odintsov
and Ying-li~Zhang for very useful discussions. E.E. was supported in
part by MICINN (Spain), projects FIS2006-02842 and FIS2010-15640, by
the CPAN Consolider Ingenio Project, and by AGAUR (Generalitat de
Ca\-ta\-lu\-nya), contract 2009SGR-994. E.P. and S.V. are supported in
part by the RFBR grant 11-01-00894, E.P. by a state
contract of the Russian Ministry of Education and Science
14.740.12.0846, and S.V. by the Russian Ministry of
Education and Science under grant NSh-4142.2010.2, and by contract CPAN10-PD12 (ICE,
Barcelona, Spain).

\end{document}